# Bayes' theorem and quantum retrodiction

STEPHEN M. BARNETT[†], DAVID T. PEGG[‡] and JOHN JEFFERS[†]

[†]Department of Physics and Applied Physics, University of Strathclyde, Glasgow G4 0NG, Scotland

[‡]Faculty of Science, Griffith University, Nathan, Brisbane 4111, Australia

**Abstract.** We derive on the basis of Bayes' theorem a simple but general expression for the retrodicted premeasurement state associated with the result of any measurement. The retrodictive density operator is the normalised probability operator measure element associated with the result. We examine applications to quantum optical cryptography and to the optical beam splitter.

# 1. Introduction

Prediction is the determination of future events on the basis of present knowledge. If our knowledge is insufficient to allow a precise determination of future events, then we must be satisfied with predicting probabilities. Often we attempt to determine past events on the basis of present knowledge. A simple example is the reconstruction of a message from a received signal. Again, if our knowledge is insufficient to allow a precise determination of past events, we must be satisfied with *retrodicting* probabilities. In our communications example, this would arise if the signal were corrupted by a noisy channel [1].

In quantum mechanics there is always an element of probability. In the predictive formulation, even if we know the state of a system precisely before measurement and have full knowledge of the measurement device, the outcome of measurement cannot, in general, be determined with certainty. It is customary, following von Neumann [2], to think of the evolution of the system in terms of a deterministic, time-reversible, unitary part and a sudden probabilistic change upon measurement. In the predictive formulation we assign the premeasurement state of the system by prediction on the basis of the outcome of some previous preparation event. In the retrodictive formalism [3-6] the state of the system between preparation and measurement is assigned retrodictively on the basis of the outcome of the measurement. The sudden change occurs at the time of preparation, so the retrodictive formalism is not simply the time reversal of the predictive formalism. Careful use of the retrodictive formalism will yield the same observable correlations as those from the predictive formalism, so the fact that the two formalisms ascribe different states to the system between preparation and measurement merely emphasises that the state is no more than a mathematical convenience embodying knowledge about the system rather than having a physical existence in its own right.

In the predictive formalism, the most general description of a measurement involves a probability operator measure (POM) [7]. In the retrodictive formalism, we can similarly describe the preparation in terms of a preparation POM. In this paper we use Bayes' theorem [8] to show how the retrodictive premeasurement density operator is simply related to the associated measurement POM element and how the (more usual) predictive density operator is similarly related to the preparation POM. We concentrate on situations in which the system does not evolve between preparation and measurement use the relation derived in some quantum optical applications.

## 2. Bayes' theorem

It is well known that probabilities are not objective but depend on the amount of knowledge available. Consider, for example, picking the winner in a ten-horse race. With no further knowledge, the probability of any particular horse winning is 1/10. Given the *extra* knowledge that one horse has been scratched, the probability of any one of the remaining horses winning is 1/9. An unscrupulous person who has fixed the race might be able to predict the winner with (near) unit probability based on this privileged *additional* information. Bayes' theorem [8] determines how additional knowledge changes the probabilities from an *a priori* distribution to an *a posteriori* distribution. It is important not to attach any temporal significance to the terms *a priori* and *a posteriori*: they simply mean with less knowledge and with more knowledge respectively. Prediction and retrodiction, on the other hand, do have temporal meanings. Prediction is determination of probabilities of later events given probabilities of earlier events. Retrodiction is determination of probabilities of earlier events given probabilities of later events. Bayes' theorem gives a means of predicting and retrodicting and also of connecting the two.

Consider two sets of possible events $\{a_i\}$ and $\{b_j\}$. Bayes' theorem states that the probability of two particular events $a_i$ and $b_j$ both occurring is

$$P(a_i, b_j) = P(a_i|b_j)P(b_j) = P(b_j|a_i)P(a_i) \,, \tag{1}$$

where $P(x|y)$ means the conditional probability of $x$ occurring given that $y$ occurs. Let the event $a$ precede the event $b$. Then the predictive and retrodictive conditional probabilities are given respectively by

$$P(b_j|a_i) = \frac{P(a_i,b_j)}{P(a_i)} , \qquad (2)$$

$$P(a_i|b_j) = \frac{P(a_i,b_j)}{P(b_j)} . \qquad (3)$$

We usually have access to the predictive conditional probability, but can use Bayes' theorem to obtain the retrodictive conditional probability as

$$P(a_i|b_j) = \frac{P(b_j|a_i)P(a_i)}{P(b_j)} = \frac{P(b_j|a_i)P(a_i)}{\sum_k P(b_j|a_k)P(a_k)} . \qquad (4)$$

In the absence of any knowledge of preference among the events $a_i$, we assign them equal *a priori* probabilities and in this case the retrodictive conditional probability is

$$P(a_i|b_j) = \frac{P(b_j|a_i)}{\sum_k P(b_j|a_k)} . \qquad (5)$$

As an example, consider the following problem. An employee makes a choice each day between going to work by bus or by train. The *a priori* probability of the bus or train being chosen is $P(B)$ or $P(T) = 1 - P(B)$ respectively. The predictive conditional probabilities of being late if the bus or train is chosen are $P(late|B)$ and $P(late|T)$ respectively. Before the employee arrives, the boss can use these probabilities to *predict* the probability that the employee will be late as

$$P(late) = P(late|B)P(B) + P(late|T)P(T) . \qquad (6)$$

When the employee indeed arrives late, the boss can then use Bayes' theorem to *retrodict* the probability that the employee has taken the bus as

$$P(B|late) = \frac{P(late|B)P(B)}{P(late)}, \quad (7)$$

where the denominator can be calculated from (6). If the boss has no knowledge of preference in the choice of transport, then he assigns a probability of 1/2 to each. This leads him to calculate the probability of his employee being late to be

$$P(late) = \tfrac{1}{2} P(late|B) + \tfrac{1}{2} P(late|T). \quad (8)$$

Then if the employee is late, the boss would retrodict the probability that the employee took the bus to be

$$P(B|late) = \frac{P(late|B)}{P(late|B) + P(late|T)}. \quad (9)$$

Naturally, the boss can change this probability by acquiring additional information for example by asking the employee which route was taken.

### 3. Application to quantum retrodiction

The usual, that is predictive, description of quantum mechanics involves a state and observables. The state is assigned on the basis of the preparation procedure. The state of the system then evovles until it is measured. In this paper, in order to highlight the roles of state preparation and measurement, we shall consider only the case where the system does not change between preparation and measurement. Two simple examples of this case are when measurement immediately follows preparation or when the Hamiltonian is zero. The observables are quantities associated with a measuring device. The probabilities associated with the results of observations are depend on measurement utilises a probability operator measure (POM) [7]. A measurement POM is characterised by its elements $\hat{\Pi}_j$. These are operators with only positive or zero eigenvalues and have the property

$$\sum_j \hat{\Pi}_j = \hat{1} ,  \qquad (10)$$

where $\hat{1}$ is the unit operator in the state space of the system. The probability that a measurement will result in the outcome $b_j$ given that the event $a_i$ that the system was prepared in the state with density operator $\hat{\rho}_i$ is

$$P(b_j|a_i) = Tr(\hat{\rho}_i^{\text{pred}} \hat{\Pi}_j) . \qquad (11)$$

For a simple von Neumann measurement, the POM elements become projectors onto the eigenstates of the operator corresponding to the measured observable so that this probability becomes the expectation value of the density operator in the associated eigenstate.

Our aim is to develop a retrodictive picture corresponding to the above. Let $P(a_i)$ be the probability that the event $a_i$ occurred (corresponding in the predictive picture to the preparation of the state $\hat{\rho}_i^{\text{pred}}$). We say that the preparation device is *unbiased* if

$$\sum_i P(a_i) \hat{\rho}_i^{\text{pred}} = \frac{1}{D} \hat{1} , \qquad (12)$$

where $D$ is the dimension of the state space[†]. We use the concept of an unbiased source to simplify our derivation of the retrodictive density operator and to illustrate more directly the connection with the predictive formalism. Use of an unbiased source corresponds to no *a priori* information about the source. This ensures that the

---

[†] If we wish to work with an infinite dimensional space then an appropriate limiting procedure must be

retrodictive state inferred on the basis of a future measurement can be based only on the result of that measurement. It corresponds to the situation in the predictive picture in which we assign states purely on the basis of earlier preparation events. Equation (12) allows us to define a preparation POM with elements

$$\hat{\Xi}_i = DP(a_i)\hat{\rho}_i^{\text{pred}} \tag{13}$$

such that

$$\sum_i \hat{\Xi}_i = \hat{1} \ . \tag{14}$$

The question now is whether we can derive a retrodictive density operator $\hat{\rho}_j^{\text{retr}}$ such that the probability that the event $a_i$ occurred, given that the later measurement yields the result $b_j$, is

$$P(a_i|b_j) = Tr(\hat{\rho}_j^{\text{retr}} \hat{\Xi}_i) \ , \tag{15}$$

for all possible preparation POM elements $\hat{\Xi}_i$.

We can use Bayes' theorem to determine $\hat{\rho}_j^{\text{retr}}$. From (1) we have

$$P(a_i|b_j) = \frac{P(b_j|a_i)P(a_i)}{P(b_j)} \ . \tag{16}$$

Substituting (11) and (15) for the predictive and retrodictive conditional probabilities we find

$$Tr(\hat{\rho}_j^{\text{retr}} \hat{\Xi}_i) = \frac{Tr(\hat{\Pi}_j \hat{\rho}_i^{\text{pred}})P(a_i)}{P(b_j)} = \frac{Tr(\hat{\Xi}_i \hat{\Pi}_j)}{DP(b_j)} \ , \tag{17}$$

where we have also used the definition (13). The *a priori* probability of the outcome $b_j$ is

$$P(b_j) = \sum_i P(b_j|a_i)P(a_i) = \sum_i P(a_i)Tr(\hat{\rho}_i^{pred}\hat{\Pi}_j) = \frac{1}{D}Tr(\hat{\Pi}_j) \qquad (18)$$

from (12). Note that this is simply the expectation value of the POM element $\hat{\Pi}_j$ for the zero information state corresponding to the density operator $D^{-1}\hat{1}$.

$$Tr(\hat{\rho}_j^{retr}\hat{\Xi}_i) = \frac{Tr(\hat{\Pi}_j\hat{\Xi}_i)}{Tr(\hat{\Pi}_j)} \quad . \qquad (19)$$

We require this relationship to hold for all possible $\hat{\Xi}_i$ corresponding to all possible preparation POMs and hence we must set

$$\hat{\rho}_j^{retr} = \frac{\hat{\Pi}_j}{Tr(\hat{\Pi}_j)} \qquad (20)$$

is the retrodictive density operator associated with the measurement outcome $b_j$. This is the main result of our paper. An analogous relationship between the predictive density operator and the preparation POM element follows from our definition (13).

We conclude this section by examining the retrodictive picture for biased sources, that is sources which produce states with predictive density operators $\hat{\rho}_{Bi}^{pred}$ prepared with *a priori* probabilities $P_B(a_i)$ which do not satisfy equation (12). We introduce operators

$$\hat{\Lambda}_i = P_B(a_i)\hat{\rho}_{Bi}^{pred} \qquad (21)$$

analogous to the predictive POM elements, but with

$$\sum_i \hat{\Lambda}_i \neq D\hat{1} \ . \tag{22}$$

We can use Bayes' theorem and the predictive conditional probability (11) to write

$$P(a_i|b_j) = \frac{Tr(\hat{\rho}_{Bi}^{\text{pred}} \hat{\Pi}_j) P_B(a_i)}{\sum_k Tr(\hat{\rho}_{Bk}^{\text{pred}} \hat{\Pi}_j) P_B(a_k)} = \frac{Tr(\hat{\Lambda}_i \hat{\Pi}_j)}{\sum_k Tr(\hat{\Lambda}_k \hat{\Pi}_j)} \ . \tag{23}$$

We can rewrite this in the retrodictive picture using the retrodictive state (20) as

$$P(a_i|b_j) = \frac{Tr(\hat{\rho}_j^{\text{retr}} \hat{\Lambda}_i)}{\sum_k Tr(\hat{\rho}_j^{\text{retr}} \hat{\Lambda}_k)} \ . \tag{24}$$

The lack of symmetry between this retrodictive conditional probability and the predictive conditional probability (11) does not reflect any intrinsic time asymmetry arising from quantum mechanics or from Bayes' theorem. Rather, it reflects the fact that it is possible, and even usual, to use biased sources of states in experiments. It is also possible, however, to make predictions which are biased on the basis of future measurements. In such cases, we would restrict our attention to only a subset of the possible experimental results. The predictive conditional probability would then be

$$P(b_j|a_i) = \frac{Tr(\hat{\rho}_i^{\text{pred}} \hat{\Pi}_j)}{\sum_l Tr(\hat{\rho}_i^{\text{pred}} \hat{\Pi}_l)} \ , \tag{25}$$

where the sum includes only those measurement POM elements in the subset of interest. This situation is not uncommon in quantum optics. Consider, for example, the much studied phenomenon of two-photon interference in which two photons from a parametric down-conversion source exhibit a non-classical effect [9]. The experiment is usually analysed with the aid of a state involving only two-photons. The theory of the parametric down-converter, however, predicts a state which is a superposition of this state and the vacuum plus other higher photon-number states. In considering only the two-photon component of the state we are restricting our description to those situations in which the future measurements record some photocounts.

## 4. Quantum optical examples

The retrodictive formalism can be used to analyse situations involving preparation and measurement for a quantum system. It will provide a more natural or simpler description of events in some situations than in others. The retrodictive interpretation of phenomena may be quite different from the more familiar predictive interpretation but is no less valid. We have already described the retrodictive interpretation of photon antibunching and of the Kocher-Commins experiment [4]. Here, we derive the retrodictive states arising in quantum cryptography and optical fields superposed by a beam splitter.

### 4.1. *Quantum cryptography*

A very important and obvious application of retrodiction is communication. The recipient of the signal (Bob) has the task of constructing the original message sent by the transmitter (Alice). In order to do this, Bob uses the knowledge gained from his measurement together with the characteristics of the communication channel. The most developed technique in quantum communications is quantum cryptography or quantum key distribution [10], in which Alice and Bob attempt to construct a secret shared key which has not been leaked to any eavesdropper.

The first protocol for quantum key distribution was devised by Bennett and Brassard [11]. In the Bennett-Brassard protocol Alice prepares single photons in either in one of two states of linear polarisation (vertical $|V\rangle$ or horizontal $|H\rangle$) or in one of two states of circular polarisation (left $|L\rangle$ or right $|R\rangle$). These states do not change on propagation to Bob. The states of circular polarisation can be written as superpositions of the states of linear polarisation and vice versa

$$|L\rangle = 2^{-\frac{1}{2}}(|V\rangle + i|H\rangle), \qquad |R\rangle = 2^{-\frac{1}{2}}(|V\rangle - i|H\rangle)$$

$$|V\rangle = 2^{-\frac{1}{2}}(|L\rangle + |R\rangle), \qquad |H\rangle = 2^{-\frac{1}{2}}i(|R\rangle - |L\rangle) \ . \tag{26}$$

Alice sets her preparation device so that in any given time slot of these states is prepared. In order to ensure security, she selects the states with equal probabilities thereby providing an unbiased source. Bob chooses to measure either circular or linear polarisation for each time slot. He chooses randomly between these two incompatible observables. Let us suppose that Alice chooses $L$ for one particular photon. On the basis of this choice, Alice assigns to the photon the *predictive* state $|L\rangle^{\text{pred}}$. She can then use this to predict the probabilities for the outcomes of Bob's measurements conditioned on her knowledge of the preparation. Alice's predictive conditional probabilities that Bob will find each of the four possible outcomes are

$$P(b_L|a_L) = P(\text{circular measured}) \times |\langle L|L\rangle^{\text{pred}}|^2 = \tfrac{1}{2} \times 1 = \tfrac{1}{2}$$

$$P(b_R|a_L) = P(\text{circular measured}) \times |\langle R|L\rangle^{\text{pred}}|^2 = \tfrac{1}{2} \times 0 = 0$$

$$P(b_V|a_L) = P(\text{linear measured}) \times |\langle V|L\rangle^{\text{pred}}|^2 = \tfrac{1}{2} \times \tfrac{1}{2} = \tfrac{1}{4}$$

$$P(b_H|a_L) = P(\text{linear measured}) \times |\langle H|L\rangle^{\text{pred}}|^2 = \tfrac{1}{2} \times \tfrac{1}{2} = \tfrac{1}{4} \ . \tag{27}$$

Let us suppose that Bob measures linear polarisation in a given time slot and finds vertical polarisation corresponding to the POM element $|V\rangle\langle V|$. From (20) he assigns to the photon the corresponding retrodictive density operator $\hat{\rho}_V^{\text{retr}} = |V\rangle\langle V|$ or equivalently the state vector $|V\rangle^{\text{retr}}$. He can then use this to retrodict the probabilities for Alice's preparation events conditioned on his knowledge of the measurement. Bob's

retrodictive conditional probabilities that Alice prepared each of the four possible states are

$$P(a_L|b_V) = P(\text{circular prepared}) \times |\langle L|V\rangle^{\text{retr}}|^2 = \tfrac{1}{2} \times \tfrac{1}{2} = \tfrac{1}{4}$$

$$P(a_R|b_V) = P(\text{circular prepared}) \times |\langle R|V\rangle^{\text{retr}}|^2 = \tfrac{1}{2} \times \tfrac{1}{2} = \tfrac{1}{4}$$

$$P(a_V|b_V) = P(\text{linear prepared}) \times |\langle V|V\rangle^{\text{retr}}|^2 = \tfrac{1}{2} \times 1 = \tfrac{1}{2}$$

$$P(a_H|b_V) = P(\text{linear prepared}) \times |\langle H|V\rangle^{\text{retr}}|^2 = \tfrac{1}{2} \times 0 = 0 \ . \tag{28}$$

It is straightforward to verify that the two sets of conditional probabilities satisfy Bayes' theorem (1) with $P(a_i) = \tfrac{1}{4} = P(b_j)$. The protocol involves a public discussion between Alice and Bob in which some of the results are discussed. Eavesdropper activity is revealed by comparing the above conditional probabilities with the observed correlations. Alternatively eavesdropper activity is indicated if, in any time slot, Alice's predictive state is orthogonal to Bob's retrodictive state.

4.2. *Beam splitters*

We shall now study the retrodictive description of the optical measuring device depicted in Figure 1. This consists of two photodetectors $D_b$ and $D_c$ in the output modes $b$ and $c$ of a beam splitter with a known (predictive) state $\hat{\rho}_c$ in the input mode $c$. The separate POM elements for the detectors are $\hat{\Pi}_n^b$ and $\hat{\Pi}_m^c$ corresponding to detection events $n$ and $m$.

The POM element for the system comprising the two detectors is $\hat{\Pi}_n^b \hat{\Pi}_m^c$. To include the beam splitter as well we change this to $\hat{U}^\dagger \hat{\Pi}_n^b \hat{\Pi}_m^c \hat{U}$ where $\hat{U}$ is the forward unitary evolution operator for the beam splitter given in terms of the creation and annihilation operators for the modes by [12]

$$\hat{U} = \exp\left[i\theta\left(\hat{b}^{\dagger}\hat{c} + \hat{c}^{\dagger}\hat{b}\right)\right] . \tag{29}$$

The reason for inserting the forward and backward unitary operators in the order we have done will become clear below. Finally we wish to find a POM element which allows us to include the field in state $\hat{\rho}_c$ as part of the complete measuring device. We require a POM element $\hat{\Pi}_{nm}$ such that when there is an incoming (predictive) state $\hat{\rho}_b^{\text{pred}}$ in mode $b$ we obtain the probability for events $n$ and $m$

$$Tr_b\left(\hat{\rho}_b^{\text{pred}}\hat{\Pi}_{nm}\right) = Tr_{bc}\left(\hat{\rho}_b^{\text{pred}}\hat{\rho}_c\, \hat{U}^{\dagger}\hat{\Pi}_n^b\, \hat{\Pi}_m^c\hat{U}\right)$$

$$= Tr_{bc}\left(\hat{U}\hat{\rho}_b^{\text{pred}}\hat{\rho}_c\, \hat{U}^{\dagger}\hat{\Pi}_n^b\, \hat{\Pi}_m^c\right)., \tag{30}$$

where we have used the cyclic property of the trace. The reason for this requirement is that the second line of (30) is the probability for these events calculated in the standard predictive formalism using the forward evolved predictive density matrix $\hat{U}\hat{\rho}_b^{\text{pred}}\hat{\rho}_c\, \hat{U}^{\dagger}$. For equation (30) to hold for any $\hat{\rho}_b^{\text{pred}}$ we require

$$\hat{\Pi}_{nm} = Tr_c\left(\hat{\rho}_c\hat{U}^{\dagger}\hat{\Pi}_n^b\, \hat{\Pi}_m^c\hat{U}\right) . \tag{31}$$

From our relationship (20) the retrodictive state of the field in input mode $b$ immediately before measurement, based on the measurement outcome of events $n$ and $m$, is

$$\hat{\rho}_b^{\text{retr}} = \frac{Tr_c\left(\hat{\rho}_c\hat{U}^{\dagger}\hat{\Pi}_n^b\, \hat{\Pi}_m^c\hat{U}\right)}{Tr_{bc}\left(\hat{\rho}_c\hat{U}^{\dagger}\hat{\Pi}_n^b\, \hat{\Pi}_m^c\hat{U}\right)} . \tag{32}$$

We shall now consider two important special cases of the above. The first involves modelling an inefficient photodetector by a perfectly efficient detector $D_b$ and a completely inefficient detector $D_c$. The state in input mode $c$ is the vacuum, $|0\rangle_{c\,c}\langle 0|$. The individual POM elements for the detectors $D_b$ and $D_c$ are respectively $|n\rangle_{b\,b}\langle n|$ and $\hat{1}_c$ where the latter is the unit operator for the state space of mode $c$. These POM elements are associated with recording $n$ photocounts in $D_b$ and the inevitable event that zero counts are recorded in $D_c$ respectively. The retrodictive state (32) now becomes

$$\hat{\rho}_b^{retr} = \frac{{}_c\langle 0|\hat{U}^\dagger|n\rangle_b{}_b\langle n|\otimes \hat{1}_c \hat{U}|0\rangle_c}{Tr_b\left({}_c\langle 0|\hat{U}^\dagger|n\rangle_b{}_b\langle n|\otimes \hat{1}_c \hat{U}|0\rangle_c\right)} \quad . \tag{33}$$

We find a simple expression for this retrodictive density operator by using the operator identity [12]

$$:\exp\left(-\hat{b}^\dagger \hat{b}\right): = |0\rangle_b{}_b\langle 0| \tag{34}$$

to give

$${}_c\langle 0|\hat{U}^\dagger|n\rangle_b{}_b\langle n|\otimes \hat{1}_c \hat{U}|0\rangle_c = \frac{1}{n!}{}_c\langle 0|\hat{U}^\dagger\left(\hat{b}^\dagger\right)^n:\exp\left(-\hat{b}^\dagger \hat{b}\right):\hat{b}^n \hat{U}|0\rangle_c , \tag{35}$$

which upon performing the unitary transformation becomes

$${}_c\langle 0|\hat{U}^\dagger|n\rangle_b{}_b\langle n|\otimes \hat{1}_c \hat{U}|0\rangle_c = \frac{1}{n!}\eta^n\left(\hat{b}^\dagger\right)^n:\exp\left(-\eta\hat{b}^\dagger \hat{b}\right):\hat{b}^n . \tag{36}$$

Here $\eta = \cos^2 \theta$ is the efficiency of imperfect detector being modelled. The denominator in (33) is most easily evaluated as the trace of (36) in the coherent state basis which gives the result $1/\eta$. Hence the retrodictive premeasurement density operator for mode $b$, conditioned on the event that the imperfect detector recorded $n$ photocounts is

$$\hat{\rho}_b^{retr} = \frac{1}{n!}\eta^{n+1}\left(\hat{b}^\dagger\right)^n:\exp\left(-\eta\hat{b}^\dagger \hat{b}\right):\hat{b}^n . \tag{37}$$

This is in agreement with our previous work in which Bayes' theorem was applied to analyse a single imperfect photodetector [5].

Our second example is the measurement method of projection synthesis [13]. This allows us to determine the probability distribution for any physical observable associated with a quantised optical field mode. To keep the discussion simple, we will assume that both detectors are perfect so that their individual POM elements are $|n\rangle_b {}_b\langle n|$ and $|m\rangle_c {}_c\langle m|$. These correspond to the events $n$ and $m$ in which $n$ and $m$ photocounts are recorded in the detectors $D_b$ and $D_c$ respectively. We prepare the reference field in input mode $c$ to be the pure state $|c\rangle_c = c_0|0\rangle_c + c_1|1\rangle_c + \text{L}$. The retrodictive premeasurement state (32) then becomes

$$\hat{\rho}_b^{\text{retr}} = \frac{{}_c\langle c|\hat{U}^\dagger|n\rangle_b {}_b\langle n|\otimes|m\rangle_c {}_c\langle m|\hat{U}|c\rangle_c}{Tr_b\left({}_c\langle c|\hat{U}^\dagger|n\rangle_b {}_b\langle n|\otimes|m\rangle_c {}_c\langle m|\hat{U}|c\rangle_c\right)} \quad . \tag{38}$$

We can construct any retrodictive state, containing up to $n+m$ photons, by a suitable choice of the reference state. Measurement by projection synthesis involves choosing the reference state such that the retrodictive state is an eigenstate of the observable of interest. As a specific example, we let the events $n$ and $m$ be the registering of 1 and 0 photocounts respectively. From (38) we find that the retrodictive state in input mode $b$ is the pure state with density operator

$$\hat{\rho}_b^{\text{retr}} = \frac{\left(c_0^* \cos\theta|1\rangle_b - ic_1^* \sin\theta|0\rangle_b\right)\left(c_0 \cos\theta\, {}_b\langle 1| + ic_1 \sin\theta\, {}_b\langle 0|\right)}{|c_0|^2 \cos^2\theta + |c_1|^2 \sin^2\theta} \quad . \tag{39}$$

Knowledge of the retrodictive state can be used for the *preparation* of selected predictive states. This can done, for example, if the field in input mode $b$ is prepared in an entangled state with a field in some other mode $d$. Retrodicting the state in input mode $b$ from the result of the measurement and projecting this onto the entangled state, allows us to collapse the predictive state for mode $d$ into an unentangled field state. As an example, consider the entangled state

$$|\Psi\rangle = \tfrac{1}{\sqrt{2}}\left(|1\rangle_d|0\rangle_b + i|0\rangle_d|1\rangle_b\right) \qquad (40)$$

which can be produced by beam-splitting a single photon.  Projecting the retrodictive pure state (39) onto the entangled predictive state (40) results in a predictive state

$$|\psi\rangle_d = \frac{c_0 \cos\theta |0\rangle_d + c_1 \sin\theta |1\rangle_d}{\left(|c_0|^2 \cos^2\theta + |c_1|^2 \sin^2\theta\right)^{1/2}} \qquad (41)$$

for the field in mode $d$.  This is the retrodictive mechanism explaining the quantum scissors device for truncating states which we have described elsewhere [14,6].

## 5. Conclusion

We have used Bayes' theorem to establish a relationship between the predictive and retrodictive formalisms of quantum mechanics.  In particular we have shown that the premeasurement density operator for the state retrodicted on the basis of the outcome of a measurement is just the normalised probability operator measure (POM) element for the general measuring device.  As well as providing a straightforward method for calculating the retrodictive state, it also provides a new physical interpretation of POM elements.  Similarly the density operator for a state predicted on the basis of the outcome of a preparation is the normalised preparation POM associated with that outcome.

We have illustrated the use of our general formula for quantum optical applications involving quantum cryptography and a quantum optical measuring device comprising two photodetectors in the output modes of a beam splitter with some known state in one of the two input ports.  We have been able to obtain results in agreement with those we have previously derived by less direct techniques and have indicated how the retrodictive formalism can be used, by means of entanglement, in the preparation of selected states.

In this paper we have restricted ourselves to the situation in which the system does not change between preparation and measurement. Where the evolution of the system density operator is unitary it is straightforward to modify our derivation of the retrodictive density operator to include the appropriate time evolution operator. For open systems, in which the evolution involves a coupling to an environment, the derivation is more involved. We address this problem elsewhere [15]. At a more fundamental level, future work might give more consideration to the role of preparation POMs, or preparables, which express the properties with which a system can be prepared, as opposed to measurement POMs, or observables, which express the properties which can be observed or measured.

**Acknowledgments**

We thank Howard Wiseman, Ottavia Jedrkiewicz and Rodney Loudon for helpful discussions and David Maguire for a careful reading of the manuscript. This work was supported by the U.K. Engineering and Physical Sciences Research Council, the Australian Research Council and the Royal Society.

**Figure Caption**

Figure 1. Optical measuring device comprising a beam splitter and photodetectors in the two output modes *b* and *c*. A field in a known state is in input mode *c*.

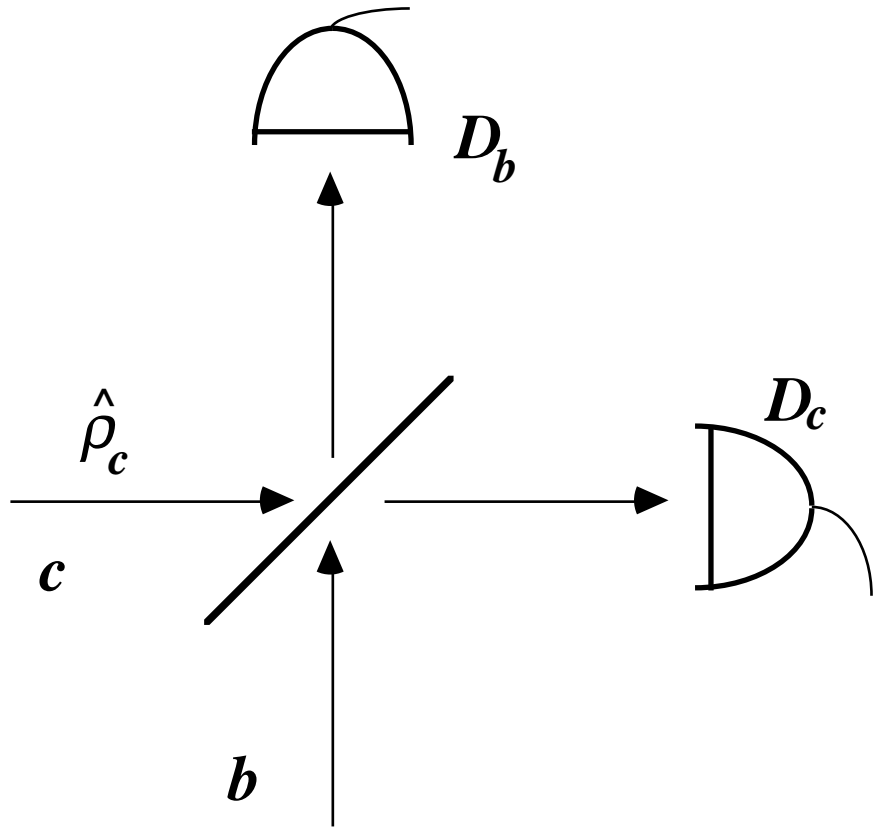

Fig. 1